\documentclass[useAMS]{mn2e}
\usepackage{lscape}
\usepackage{rotating}
\usepackage{amssymb}
\usepackage{color}
\def\Msun{\hbox{M$_\odot$}}

\def\msun{\hbox{M$_\odot$}}

\def\t4{\hbox{t$_{\rm 4}$}}

\def\cm3{\hbox{cm$^{-3}$}}
\input psfig.sty
\input epsf.sty
\usepackage{graphicx}
\usepackage{epsfig}
\title[Age Dating in the Near Infrared]
{Age Dating Stellar Populations in the Near Infrared: An absolute age indicator from the presence/absence of red supergiants}
\author[Gazak et al.] {J.~Z. Gazak$^{1}$, N. Bastian$^{2,3}$, R.-P. Kudritzki$^{1,4,5}$, A. Adamo$^{6}$, B. Davies$^{3,7,8}$ \newauthor B. Plez$^9$, M.~A. Urbaneja$^1$ \\
$^1$ Institute for Astronomy, University of Hawaii, 2680 Woodlawn Drive, Honolulu, HI 96822, USA\\
$^2$ Excellence Cluster Universe, Boltzmannstr. 2, 85748 Garching, Germany\\
$^3$ Astrophysics Research Institute, Liverpool John Moores University, Egerton Wharf, Birkenhead, CH41 1LD, UK\\
$^4$ Max-Planck-Institute for Astrophysics, Karl-Schwarzschild-Str.1, D-85741 Garching, Germany\\
$^5$ University Observatory Munich, Scheinerstr. 1, D-81679 Munich, Germany\\
$^6$ Max-Planck-Institut for Astronomy, K\"onigstuhl 17, D-69117 Heidelberg, Germany\\
$^7$ Institute of Astronomy, University of Cambridge, Madingley Road, Cambridge CB3 0HA, UK\\
$^8$ School of Physics \& Astronomy, University of Leeds, Woodhouse Lane, Leeds LS2 9JT, UK\\
$^9$ Laboratoire Univers et Particules de Montpellier, Universit\'e Montpellier 2, CNRS, F-34095 Montpellier, France
}
\date{Accepted for publication in MNRAS, November 2012.}
\pagerange{\pageref{firstpage}--\pageref{lastpage}}
\pubyear{2012}
\begin{document}
\maketitle

\label{firstpage}
\begin{abstract}
The determination of age is a critical component in the study of a population of stellar clusters. In this letter we present a new absolute age indicator for young massive star clusters based on J-H colour. This novel method identifies clusters as older or younger than 5.7 $\pm$ 0.8 Myr based on the appearance of the first population of red supergiant stars. We test the technique on the stellar cluster population of the nearby spiral galaxy, M83, finding good agreement with the theoretical predictions. The localisation of this technique to the near-IR promises that it may be used well into the future with space$-$ and ground$-$based missions optimised for near-IR observations.

\end{abstract}

\section{Introduction}
\label{sec:intro}

In order to study a population of stellar clusters, i.e. to derive cluster dispersal times, relate cluster to star-formation events in the galaxy, or estimate their impact on the surrounding interstellar medium, the ages of individual clusters must be accurately measured.  For this purpose, there have been a variety of techniques proposed in the literature to estimate the age of young massive clusters (YMCs), both in the Galaxy as well as in more distant (e.g., starburst) systems.  

The techniques can be broadly split into four categories. The first compares the integrated broad/narrow band photometry (e.g., Anders et al.~2004), integrated spectral shape (e.g., Smith et al.~2006), or spectral line strengths (e.g., Trancho et al.~2007) to a grid of models (age, extinction, metallicity) that follow the evolution of the spectral energy distribution of a simple stellar population.  The second uses resolved photometry of individual stars in order to make colour-magnitude diagrams (as a proxy for the more physical Hertzsprung$-$Russell diagram) in order to compare with theoretical stellar evolutionary isochrones.  This technique has a long history in clusters within the Milky Way (see, for example, Russell 1914) and has recently been extended to massive $>10^5$\msun\ YMCs in nearby galaxies (e.g., Larsen et al.~2011).  The third category is to use some feature of the YMC, such as surface brightness fluctuations, the size of the bubble created by the YMC in the surrounding ISM (e.g, Whitmore et al. 2011), or the equivalent width of recombination lines (e.g. Leitherer \& Heckman 1995) as a proxy for the cluster age.  The fourth and final broad category is the use of specific spectral features that are due to a unique stellar evolutionary phase with known beginning time and duration.

While each technique requires accurate models of stellar evolution and spectral synthesis$-$of which our understanding and techniques continue to develop$-$the main difference is observational.  Techniques requiring only broadband photometry are observationally efficient, but spectra provide richer sources of information and may warrant the increased exposure time based on the goals of the specific project, i.e. age, metallicity, morphology, etc.  

An example of the fourth category of age dating techniques is the use of Wolf-Rayet spectral features (namely the 4650\AA\ emission ``bump") in the integrated spectra of young clusters (e.g., Bastian et al.~2006; Sidoli et al.~2006).  This type of technique is particularly attractive, as it avoids the complexities of modelling a full stellar population, and focusses on a single stellar evolutionary phase.  Here, we introduce a new technique along similar lines, based on the evolution of massive stars.  We apply this technique to a photometric catalogue of young massive clusters in M83.

Red supergiants (RSGs) are young, massive, luminous stars evolving from blue supergiants which rapidly exhaust their core hydrogen.  The tremendous luminosities of RSGs peak at $\sim$ 1$\mu$m with absolute J-band magnitudes of M$_J$ = $-8$ to $-11$, rivalling the integrated light of globular clusters and dwarf galaxies.

The near-IR flux of RSGs dominates the integrated properties of clusters which contain them.  For example, a single RSG of 25 \Msun\ at  4000 K emits 25\% of the J band flux of a 10$^5$ \Msun\ cluster when initially formed.  When the massive stars in such a cluster begin to evolve to the RSG evolutionary phase they emit in excess of 95\% of the J band light of the full population.   In this letter we discuss a new technique using near-IR colours as an absolute age indicator for young star clusters.  This technique utilizes the initial appearance of RSGs and their strong effect on the integrated properties of their host populations.  

In \S2 we describe our observational data, in \S3 we present the results of our age diagnostic simulations, and in \S4 we offer brief conclusions and remark on future applications of this and similar near-IR techniques.  Cluster simulations in this letter utilize a Saltpeter initial mass function with mass boundaries of 0.8-100 \Msun

\section{Observations}
\label{sec:obs}

\subsection{Cluster catalogue}

The catalogue of clusters in M83 was taken from Bastian et al.~(2011, 2012).  We refer the reader to those papers for details on the cluster selection, photometry and derivation of cluster properties (ages, masses, extinctions, sizes).  Here we briefly outline the methodology adopted.  Cluster candidates were selected through visual inspection of {\em Hubble Space Telescope} (HST) Wide Field Camera 3 (WFC3) V-band images, from a larger list of candidates selected through the automated {\em SExtractor} (Bertin \& Arnouts 1996) routine, to be resolved, centrally concentrated and symmetric in the inner regions of the cluster profile.  The cluster properties were derived through comparison of each clusters U (F336W), B (F438W), V (F555W or F547M), H$\alpha$ (F657N), and I-band (F814W) magnitudes with simple stellar population models (SSP) using the method described in Adamo et al. (2010a).  For the SSP models we adopted those of Zachrisson et al. (2011) that include nebular emission, and we adopted a metallicity of 2.5 times solar (Bresolin \& Kennicutt 2002, Bresolin et al. 2005).  

We emphasise that only optical and UV colours were used in deriving the age of each cluster.  Additionally, we only included cluster candidates that had inferred photometric masses above $5 \times 10^3 $\msun .  This cut was made to limit the effects of a stochastically sampled stellar initial mass function within the clusters. Catalogue clusters near the minimum mass may be affected slightly.  In clusters of lower mass, the population of massive stars poorly samples the full probability density distribution of stellar masses such that the appearance of RSGs is skewed to older ages by gaps in the cluster population.

\subsection{HST WFC3 near-IR photometry}

In the present work we also make use of the WFC3 Early Release Science images of M83 taken in the F110W (J) and F160W (H) bands.  We carried out aperture photometry on these near-IR images, using the cluster catalogue described above, with an aperture radius, background radius and annulus size of 2.5, 4, and 1 pixel, respectively, the same (in arcsec) as was used for measuring the optical colours.  The zeropoints were taken from the WFC3 HST Zeropoint website.  Aperture corrections were derived from bright and isolated sources (clusters and individual$-$likely foreground$-$stars).  We applied aperture corrections of 0.53 and 0.59 magnitudes in the F110W and F160W bands.  We note that the individual aperture correction for each sources varied, depending on the extent of which the source was resolved.  However, the difference between the F110W and F160W bands was always 0.06 (with scatter less than 0.03 mag).  Finally, we only included clusters in the sample discussed here that have photometric errors less than 0.1 mag in both near-IR bands.

\section{Results}
\label{sec:results}

\subsection{Age vs. near-IR colour}

\begin{figure}
\includegraphics[width=8.5cm]{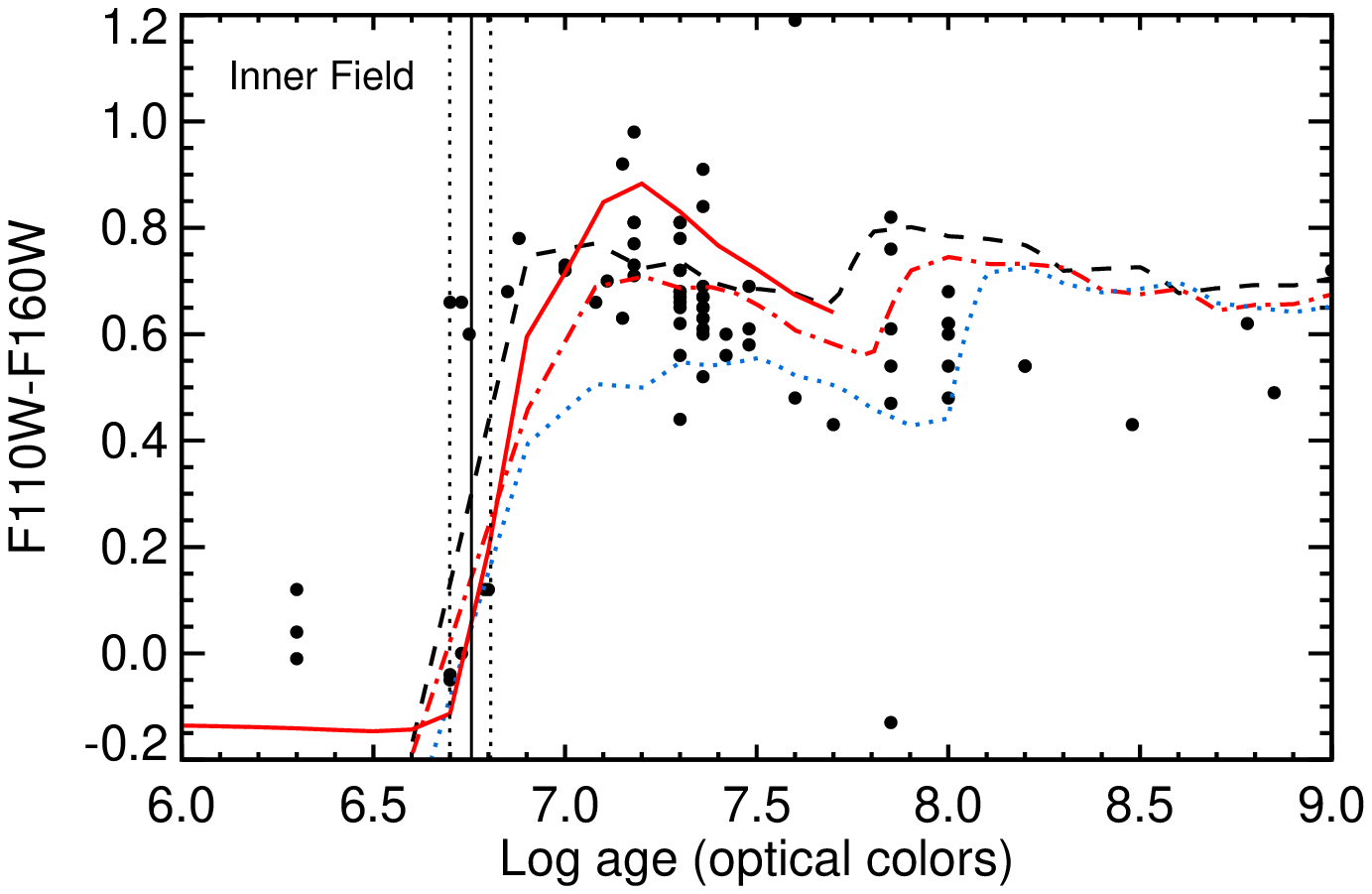}
\includegraphics[width=8.5cm]{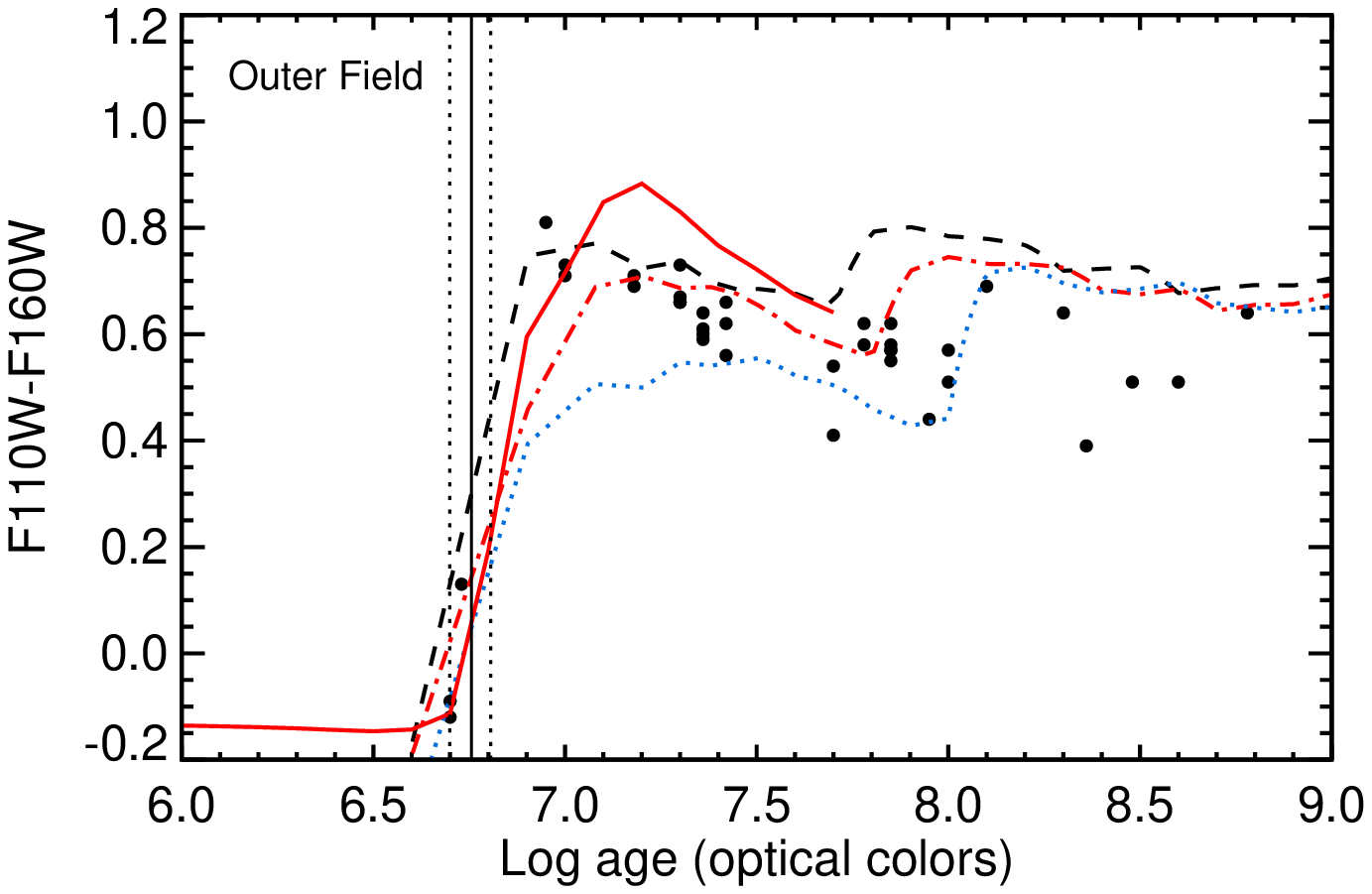}
\caption{The derived ages of the cluster sample (using optical colours) vs. the $J-H$ (F110W$-$F160W) colour for the inner field (top panel) and outer field (bottom panel) fields of M83 (Bastian et al.~2011, 2012).  These colours have not been dereddened.  The solid red curve displays our theoretically calculated colour evolution using Geneva evolutionary tracks and colours calculated from Kurucz models for the main sequence and MARCS models for RSGs.  The other curves are obtained with the GALEV SSP code for twice solar (dashed black), solar (red dashed-dotted), and half solar (blue dotted) metallicity.  Note that the GALEV simulations begin at cluster age of 6.6 [log years] while our simulations end at 7.7 [log years].  The solid and dotted vertical lines show the appearance of RSGs and uncertainty as calculated in this work.}
\label{fig:age_vs_ir}
\end{figure}

In Figure~\ref{fig:age_vs_ir} we show the F110W--F160W colour vs. the derived age (from optical colours only) for clusters in the inner field (top panel) and outer field (bottom panel).  In both fields we see a bi-modal colour distribution, with clusters with ages $\lesssim 5$~Myr having F110W--F160W colours $<0.4$, and older clusters having redder colours.  In particular, we note the appearance of a step-like change in colours at $\sim 5$~Myr.  As will be discussed below, this jump is predicted from stellar evolutionary models due to massive stars entering the RSG phase, becoming extremely luminous and very red.

The data plotted in Figure~\ref{fig:age_vs_ir} include the effects of reddening, which are relatively small in the infrared.  Using the Cardelli et al. (1989) reddening law and assuming R$_V$=3.1 we estimate $\delta$(J-H) $\sim$0.09*A$_V$.  This means that a selective increase in A$_V$ by 9 mag is needed for the older clusters exactly at the age of the J-H blue vs. red transition to produce an 0.8 mag jump in the colors.   In principle a YMC that is heavily embedded may be extincted enough to appear ``old'' in the J-H index.  However, studies of YMCs in merging galaxies have shown that the deeply embedded stage is relatively short lived, and that nearly all YMCs detected in the radio continuum are also detected in the optical, so this is unlikely to present a strong bias (Whitmore \& Zhang 2002).  The inclusion of emission line (e.g. Br$\gamma$) photometry in such an analysis would allow such deeply embedded clusters, if present, to be detected.  On the other hand, the comparison with the SSP models and simulated J-H colors (see below) is, of course, affected by the reddening.   We note that the average A$_V$ is about 0.37 and 0.30 for the cluster catalogues of the inner and outer fields, respectively.  Thus Figure~\ref{fig:age_vs_ir} is not strongly affected by reddening.  The J-H technique may be inapplicable to cases of high or strongly varied extinction, such as Arp 220 or M82.

\subsection{The effect of RSGs on near-IR cluster colours}

As a massive star evolves into a RSG following the blue supergiant phase it expands and cools while the bulk of emitted flux shifts from optical to near-IR wavelengths.  These objects evolve through the main sequence quickly$-$within 10 Myr, for masses above 15 \Msun $-$and through the blue supergiant phase even faster, in less than 1 Myr for similarly massive stars (Meynet \& Maeder 2000, Kudritzki et al. 2008).

\begin{figure}
\includegraphics[width=8.5cm]{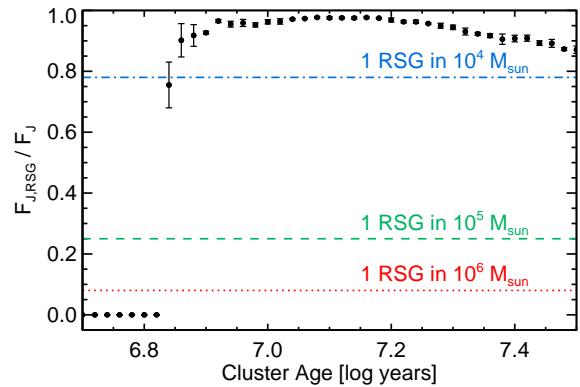}
\caption{The J-band flux contribution of RSGs in a simulated super star cluster of 10$^5$ \Msun\ using Geneva evolution tracks at solar metallicity including stellar rotation.  Error bars show estimated scatter from ten simulated clusters using a Salpeter initial mass function.  Horizontal lines represent the flux contribution from a single 25 \Msun\, 4000 K RSG injected into simulations of zero age clusters with masses of 10$^4$ \Msun\ (blue, dash-dotted), 10$^5$ \Msun\ (green, dashed), and 10$^6$ \Msun\ (red, dotted).}
\label{fig:percj}
\end{figure}

In Figure \ref{fig:percj} we show the dominance of RSGs on the J-band flux of massive star clusters as a function of age.  These simulations are done using Geneva evolutionary tracks with rotation at solar metallicity.  The work clearly shows that the appearance of RSGs marks a significant change in the observational properties of the integrated light of the cluster in the near-IR. 

The appearance of the first RSG in a young massive cluster initializes a hysteresis effect in which near-IR colour reddens significantly and remains red as increasing numbers of less luminous stars evolve to red evolutionary phases.  As a single RSG affects the near-IR colours of a massive star cluster in this way, the measurement of that change in colour represents an absolute age indicator.  

\begin{figure}
\includegraphics[width=8.5cm]{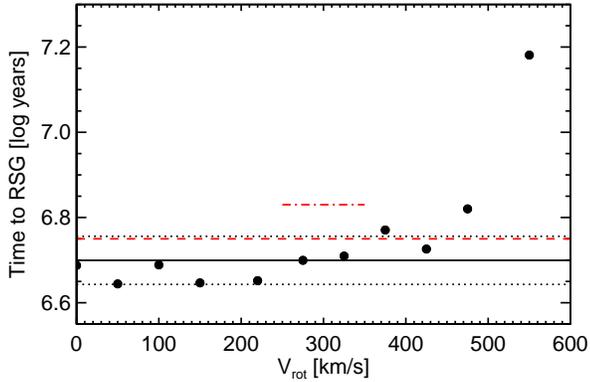}
\caption{Effects of rotation on the mass and age of the first RSGs in a simple stellar population.   Points represent the first appearance of RSGs as a function of initial rotational velocity, V$_{rot}$ (Brott et al. 2011).  The solid and dotted black lines show the average and scatter in these initial ages excluding the point at V$_{rot}$ = 550 [km/s].  The red dashed line marks the appearance of RSGs in Geneva evolution tracks without rotation, and the red dash-dotted line Geneva with rotation included at V$_{rot}$ = 0.4 V$_c$ ($\sim$ 250-350 km/s for the initial population of RSGs).  Utilizing the three sets of stellar evolution models, the initial appearance of RSGs occurs at 5.7 $\pm$ 0.8 Myr. }
\label{fig:rot_effect}
\end{figure}

Observations show that 50-60\% of massive stars form and evolve as binaries (Sana et al. 2012, Sana \& Evans 2010).  The technique presented in this letter requires that the age of the first RSGs remain stable (Figure \ref{fig:rot_effect}) and that the RSGs dominate the near-IR flux of the YMC (Figure \ref{fig:percj}).  Eldridge et al (2008) argue that, for a reasonable distribution of binary separations and mass ratios, the mean lifetime of RSGs may be reduced by a factor of 2 to 3 compared with the predictions of single-star models.  Evolution towards the RSG phase is unaffected. The most extreme effect of binarity on the stellar population is therefore to remove $\sim$60\% of the RSGs, to be replaced by blue supergiants. Though this will cause the YMC to be fainter in the near-IR, the RSGs' contribution to the cluster's total near-IR flux decreases by only a few percent, and the sharp jump in J-H colour for ages of $\ga$5Myr remains. Hence, the impact of binarity on our results is merely to increase stochasticity effects in low-mass clusters.

\begin{figure}
\includegraphics[width=8.5cm]{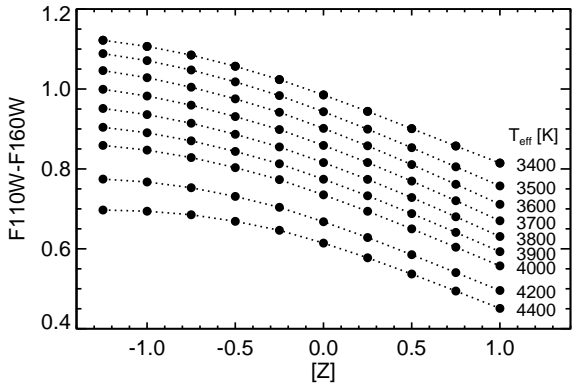}
\caption{Colour evolution of RSGs in temperature and metallicity used in this work.  Colours are measured using a new grid of {\sc marcs} models and {\sc turbospectrum} (Alvarez \& Plez 1998, Plez 2012).}
\label{fig:colvz}
\end{figure}

\subsection{Age dating clusters based on the presence/absence of RSGs}

Our simulations localize the colour break in clusters with and without RSGS to 5.7 $\pm$ 0.8 by weighting multiple techniques present in the literature to measure the timescale of formation for RSGs.  

The uncertainty inherent in the RSG age dating technique arises from any process which affects stellar evolution.  Here we describe the role of stellar rotation, which affects the maximum stellar mass which evolves to the RSG phase.  We utilize the Brott evolutionary models for this discussion as they cover a range of metallicities and rotation (Brott et al. 2011).  Figure \ref{fig:rot_effect} shows the evolution of the appearance of the first RSG in the Brott evolution tracks as a function of initial stellar rotation, as well as the appearance of RSGs in two sets of Geneva evolution tracks with and without rotation at solar metallicity (Meynet \& Maeder 2000).  

The Brott models show a slow increase in the time to the appearance of a RSG which evolves with a decreasing mass of the initial RSGs.  This effect is due to an increasing size in the hydrogen burning core of fast rotating stars.  At high rotational velocities, stars evolve as if they were more massive, and as such explode in core collapse supernova before reaching the RSG phase.  At rotation velocities above 500 km/s this effect becomes significant, lowering the mass of the first RSG to $\sim$15\Msun\ from the consistently 30-40 \Msun\ initial RSG population for less extreme rotation velocities.  Assuming a spread of velocities in a simple stellar population, we drop the 550 km/s case as all stars would not be expected to rotate with such extreme velocities and calculate an age for the appearance of RSGs to 5 $\pm$ 0.7 Myr from the Brott models.  The Geneva tracks yield slightly longer delays before RSG appearance of 5.6 Myr for tracks without rotation and 6.9 Myr for tracks with rotation.  The Geneva group models with rotation adopt initial rotation speeds of V$_{rot}$= 0.4V$_c$.  Weighting these three measurements we measure a time of RSG appearance of 5.7 $\pm$ 0.8 Myr.

As discussed in \S3.2, the effects of these initial populations of RSGs on the near-IR cluster colours are expected to appear as a significant hysteresis jump allowing for a clear differentiation between young ``blue" clusters and ``red" clusters in which massive stars have evolved into RSGs and dominate the near-IR properties.  Our cluster catalogue shows exactly such evolution in the HST J-H (F110W$-$F160W) colour as plotted in Figure \ref{fig:age_vs_ir}.  The ages of these clusters have been calculated by an independent method utilizing optical colours (see \S2), and the work in this letter shows that a clear age differentiation may be measured using only near-IR colours.   We model this effect by simulating clusters using the Geneva stellar evolution models including rotation.  Cluster colours are calculated at every time step with synthetic photometry of cluster stars, utilizing the Kurucz grid of spectral energy distributions for main sequence stars and a new grid of MARCS models  calculated for the critical RSG population (Gustafsson et al. 2008).  In Figure~\ref{fig:colvz} we plot the new grid of colours applied to RSGs in our synthetic cluster calculations.  This technique successfully recreates the form and scale of the observed effect in M83 (Figure \ref{fig:age_vs_ir}).
We recover nearly identical results using the {\sc galev} code to model our simple stellar populations over a range of metallicities from half to twice solar, and note that the effect is recovered in similar passbands using starburst99 (Kotulla et al. 2009,  Leitherer et al. 2010).

We note, however, that SSP calculations at metallicties lower than half solar with different sets of evolutionary tracks give conflicting results with regard to the steepness of the blue to red (J-H) transition. This is obviously the result of a different treatment of mass-loss and rotational mixing. A careful observational investigation of stellar clusters in low metallicity environments will be worthwhile in this regard. 

Studies such as Sharma et al. (2007) have found a significant spread in the age of the stellar population in young associations.  Naylor (2009) shows that a systematic difference in age determination between young main sequence and pre main sequence stars can explain the apparent spread in ages. Regardless, our catalogue has been specifically designed to exclude sources such as the one in Sharma et al. (2007).  For those true YMCs which remain, no evidence for such age spreads in the massive star population have been found (Kudryavtseva et al. 2012).

While near-IR colours represent an efficient method to separate young and evolved clusters, spectroscopy in the near-IR offers similar promise to constrain age.  In addition, spectroscopic observations may be used to extract abundances of Fe and $\alpha$-elements of clusters observed at even modest resolutions of R$\sim$3000 (Davies et al. 2010).   This is due to the presence of strong and well separated RSG spectral features in the near-IR and the dominance of those objects in the near-IR (Figure \ref{fig:percj}).  In two upcoming papers we discuss observed and theoretically calculated spectra to show the effects of the appearance of RSGs and the possibility of using such integrated population spectroscopy to derive metallicity information (Westmoquette et al. in prep, Gazak et al. in prep).

\section{Conclusions}

In this letter we present a new absolute age indicator for young massive star clusters using J-H colour which can be used to determine if a massive star cluster is older or younger than 5.7 $\pm$ 0.8 Myr.  Because this technique requires only near-IR photometry it can be used well beyond the HST era when both space$-$ and ground$-$based telescopes move to the near$-$IR.  The technique can be used with spectroscopic data as well and we show this in two upcoming papers.  While this age dating method is shown to work well in the range between half to twice solar metallicities, a careful observational investigation of its applicability at significantly lower metallicity is suggested, since present evolutionary tracks give conflicting results in this low metallicity domain.

\section*{Acknowledgments}

JZG and RPK acknowledge support by the National Science Foundation under grant AST-1108906 and the hospitality of the Max-Planck-Institute for Astrophysics, where part of this work was carried out.  NB is supported by the DFG cluster of excellence ÕOrigin and Structure of the UniverseÕ.  BP acknowledges support from the CNRS/INSU PNPS

\bsp
\label{lastpage}
\end{document}